\documentclass[traditabstract]{aa}
\usepackage{times}
\usepackage{natbib}
\usepackage[dvips]{graphicx}
\bibpunct{(}{)}{;}{a}{}{,}

\def\ltsima{$\; \buildrel < \over \sim \;$}
\def\lsim{\lower.5ex\hbox{\ltsima}}
\def\gtsima{$\; \buildrel > \over \sim \;$}
\def\gsim{\lower.5ex\hbox{\gtsima}}
\def\msun{\mathrm{~M_{\odot}}}
\def\rsun{\mathrm{~R_{\odot}}}

\def\sm{SMP 22}
\def\xmm{{\em XMM--Newton}}

\begin{document}

\title
{An XMM-Newton view of Planetary Nebulae in the Small Magellanic Cloud}
\subtitle
{The X-ray luminous central star of SMP SMC 22}
\author{S. Mereghetti\inst{1}, N. Krachmalnicoff\inst{1,5},  N. La Palombara\inst{1},
A. Tiengo\inst{1}, T. Rauch\inst{2}, F. Haberl\inst{3}, M.D. Filipovi\'{c}\inst{4},
R. Sturm\inst{3}}

\institute{
INAF, Istituto di Astrofisica Spaziale e Fisica
Cosmica Milano, via E.\ Bassini 15, I-20133 Milano, Italy
\and
Institute for Astronomy and Astrophysics, Kepler Center for Astro and Particle Physics, Eberhard Karls University, Sand 1, D-72076 T\"{u}bingen, Germany
\and
Max Planck Institute for Extraterrestrial Physics, D-85748 Garching, Germany
\and
University of Western Sydney, Locked Bag 1797, Penrith South DC, NSW 1797, Australia
\and
Dipartimento di Fisica, Universit\`a degli Studi di Milano,
	via G.~Celoria 16, I-20133 Milano, Italy
}
\offprints{S. Mereghetti, sandro@iasf-milano.inaf.it}

\date{Received March 15, 2010 / Accepted May 4, 2010}

\authorrunning{S. Mereghetti et al.}

\titlerunning{X-rays from the planetary nebula SMP SMC 22}

\abstract
{During an X-ray survey of the Small Magellanic Cloud, carried out with the
XMM-Newton satellite, we detected significant soft X-ray emission from the
central star of the high-excitation planetary nebula SMP SMC 22.
Its very soft spectrum is well fit with a non local thermodynamical equilibrium model atmosphere composed of H, He, C, N, and O, with abundances equal to those
inferred from studies of its nebular lines. The derived effective temperature of
1.5$\times10^5$ K is in good agreement with that found from the optical/UV data.
The unabsorbed flux in the 0.1--0.5 keV range
is $\sim3\times10^{-11}$ erg cm$^{-2}$ s$^{-1}$, corresponding to a luminosity
of $\sim1.2\times10^{37}$ erg s$^{-1}$  at the distance of 60 kpc.
We also searched for X-ray emission from a large number of SMC planetary nebulae,
confirming the previous detection of SMP SMC 25 with a luminosity
of (0.2--6)$\times10^{35}$ erg s$^{-1}$ (0.1-1 keV). For the remaining
objects that were not detected, we derived flux upper limits corresponding to
luminosity  values
from several tens to hundreds times smaller than that of SMP SMC 22.
The exceptionally high X-ray luminosity of SMP SMC 22 is probably due to the
high mass of its central star, quickly evolving toward the white dwarf's cooling
branch, and to a small intrinsic absorption in the nebula itself.}


\maketitle

\section{Introduction}

Planetary nebulae (PNe) are a common stage in the evolution of low  and intermediate mass stars,
leading to the formation of white dwarfs.
They appear when the fast wind from the central star interacts with the matter of the denser wind that was previously ejected during the asymptotic giant branch (AGB) phase,
and are characterized by  H$_{\alpha}$ line emission.
Since the advent of imaging X-ray telescopes a number of PNe have been detected in the soft X-ray range (see, e.g., \citealt{chu03}), but due to their relatively low fluxes,   detailed studies with the \xmm\ and Chandra
satellites have been carried out only for a few objects \citep{gru06,kas07,mon09}.
X-ray emission from PNe can originate from their central star or from the
hot gas shocked in the interaction between the two stellar winds. Usually one of these two
processes, characterized by different spectral and spatial signatures, is dominant, but there are also PNe in which both components have been detected.
There is also the possibility that some X-ray emission observed from PNe is
actually due to coronal emission from a companion star (see, e.g., \citealt{sok02}).

SMP SMC 22 (hereinafter \sm ) is a  high-excitation planetary nebula
located in the Small Magellanic Cloud (SMC) \citep{san78,all87} and  characterized by a very high X-ray luminosity \citep{wan91}.
Its large X-ray flux led to an  early detection of this source in the soft X-ray range
with the Einstein Observatory \citep{sew81} and to its inclusion in the class of super-soft X-ray sources (SSS), a
rather heterogeneous group of luminous (10$^{36}$--10$^{38}$ erg s$^{-1}$) sources characterized by thermal-like emission corresponding to blackbody temperatures
of $10^5$--10$^6$ K (see, e.g., \citealt{kah06}). Most SSS are binary systems
containing accreting white dwarfs, but a few of them have been identified with the nuclei
of PNe.
Here we report on recent  X-ray observations of \sm\ obtained
with the \xmm\ satellite, as well as on a systematic search for
X-ray emission from a large sample of PNe in the SMC.

\section{Observations and data analysis}
\label{obs}

The field of \sm\ has
been observed in three separate occasions with \xmm\ (see Table \ref{obs}).
The two 2009 pointings were obtained as part of our Large Program aimed at
a complete survey of the SMC \citep{hab08b}.
For completeness we also analyzed the 2007
observation available in the public archive.
The data discussed here were obtained with the EPIC instrument,
consisting of one pn and two MOS cameras \citep{str01,tur01} covering a field of
view of about 30$'$ diameter in the energy range 0.1--12 keV.
In all the observations they were operated in  full frame mode, yielding a
time resolution of 73 ms for the pn and 2.6 s for MOS1 and MOS2.
We processed the data with the standard \textit{XMM-Newton Science Analysis Software} (SAS, Version 8.0).  We filtered out time intervals affected by high background, induced by soft protons, resulting in the net exposure times listed in Table \ref{obs}.

\begin{table*}[ht!]
\centering
\begin{tabular}{ccccccc}
\noalign{\smallskip}
\hline
\noalign{\smallskip}
Id.   & Date & Duration & Net exposure & Off\,axis angle & Count rate & Tot. counts\\
                     &   &   (ks) & (ks) & (arcmin) & (cts s$^{-1}$) & \\
\noalign{\smallskip}
 \hline
 \noalign{\smallskip}
0501470101 &   2007/06/04 & 31.8 & 7.7 & 5.58  & 0.400 & 3100\\
0601210101 &   2009/05/14 & 26.7 & 16.54 & 7.20 & 0.309 & 5167 \\
0601210501 &   2009/09/25 & 48.7 & 32.3 & 14.06 & 0.153 & 5442 \\
\noalign{\smallskip}
\hline
\end{tabular}
\caption{\small{Log of the \xmm\ observations of \sm\ . All the quantities refer to the
EPIC pn camera. }}
\label{obs}
\end{table*}

To extract spectra and light curves we selected only single and double events with energy between 0.1 and 12 keV. For the first and second observation we used circular extraction regions with radii of $30''$ and $25''$, respectively. In the third observation, due to the large off-axis angle, \sm\  appears with an elongated shape; for this reason we chose an elliptical region to extract the source events. In all cases the background spectrum was obtained from events extracted in a large area with no sources in the same CCD containing \sm .
We generated the response matrices (\textit{rmf}) and the ancillary files (\textit{arf}) using the \textit{SAS} tasks \textsc{rmfgen} and \textsc{arfgen}. All source spectra extracted from the event lists were rebinned in order to have a minimum of 30 counts per energy bin and fitted using the XSPEC program.
A systematic error of 5\% was included in the spectral fits to account for the known uncertainties\footnote{EPIC status
of calibration and data analysis, http://xmm2.esac.esa.int/docs/documents/CAL-TN-0018.pdf} in the response matrices below 0.5 keV.

The light curves extracted for \sm\ are consistent with a constant flux in each
data set and the source count rates  in the three observations, when corrected for the different off-axis angle, do not show any evidence for long term variability.

The two MOS together provide
only about 10\% of the total counts detected from the source and the addition of their spectra does not significantly improve the results. Therefore,
in the following we report only the results obtained with the pn.
The spectra from the three observations were fitted simultaneously, forcing common parameters (except for the relative normalization, whose values
resulted always within 1\%).

\begin{figure}[htbp]
\centering
\includegraphics[width=6 cm, angle=-90]{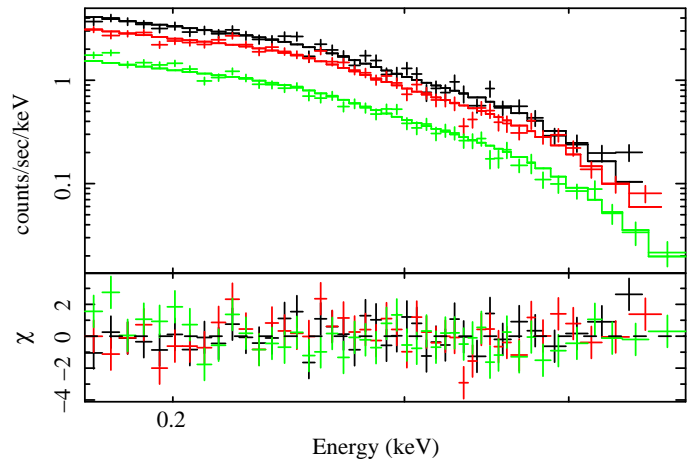}
\caption{\small{Spectra of \sm\ fitted  with a  blackbody model.
Top panel: data points and best fit model for the three observations (in chronological
order from the highest to lowest count rate). Bottom panel: residuals from the
best fit model in units of $\sigma$.}}
\label{bbody}
\end{figure}

\begin{figure}[htbp]
\centering
\includegraphics[width=6 cm, angle=-90]{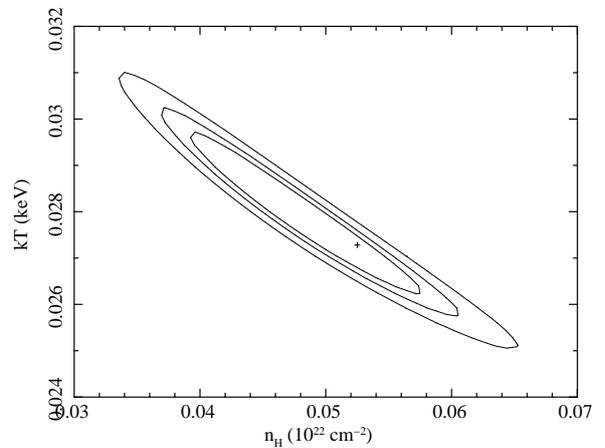}
\caption{\small{Confidence contours (68\%, 90\%, and 99\% c.l.) for the absorption and temperature
obtained with the  blackbody fit to the three joint observations.}}
\label{bbody_c}
\end{figure}

The results obtained with a few single component models confirm that the source spectrum
is very soft (see Table \ref{simple_model}). In fact, practically no source counts
were detected above 0.5 keV.
The blackbody and the
thermal plasma model (Mekal in XSPEC)  provide the best fits, with temperatures of the order of $\sim$(3--5)$\times10^5$ K
and absorption N$_{H}\sim5\times10^{20}$ cm$^{-2}$.
A power law, besides giving an unacceptable  $\chi^2$, yields an unrealistic photon index. The best fit blackbody model is shown in Fig. \ref{bbody}, and the corresponding confidence contours of
temperature and absorption in Fig. \ref{bbody_c}. At a distance of 60 kpc (adopted hereinafter; see, e.g., \citealt{deb10}),
the blackbody emitting
surface corresponds to a star radius of $0.05 \rsun$ and the bolometric luminosity is $\sim8\times10^{37}$ erg s$^{-1}$.

\begin{table*}[ht!]
\centering
\begin{tabular}{lccccccccc}
\noalign{\smallskip}
\hline
\noalign{\smallskip}
Mod. &$\chi^{2}_{red}$/dof &N$_H$ & $T$ & R &$\alpha$& Z  & $F_{X}$ & $F_{bol}$\\
           &  & ($10^{20}\, \mbox{cm}^{-2}$) & ($10^{5}$ K) & ($10^{9}$ cm) & &&($10^{-10}$ $\frac{erg}{cm^{2} s}$) &($10^{-10}$ $\frac{erg}{cm^{2} s})$\\
\noalign{\smallskip} \hline \noalign{\smallskip}
Blackbody & 0.98/124 &$5.2^{+1.3}_{-1.9}$ & $3.13^{+0.46}_{-0.23}$ & $3.4^{+3.4}_{-2.1}$ & - & - & $0.9$ & $1.9$ \\\noalign{\smallskip}
Brems. & 1.10/124 &  $4.1^{+5.2}_{-1.0}$ & $4.87\pm0.46$ &  - & - & -& $0.6$ & $12.3$\\\noalign{\smallskip}
Mekal &  0.93/123 & $5.5^{+2.0}_{-1.7}$ & $3.94\pm0.69$ & - & -& $0.02_{-0.01}^{+0.02}$  & $3.0$ & $86.1$\\\noalign{\smallskip}
Pow. law & 1.69/124 & $\geq4.89$ &- &-& $\geq9.4$&-& $\geq22.6$ & -\\
\noalign{\smallskip}
 \hline
\end{tabular}
\caption{\small{Best fit parameters and 90\% c.l. errors;  $F_{X}$ is the unabsorbed flux in the 0.1-0.5 keV energy band.}}
\label{simple_model}
\end{table*}

\begin{figure}[htbp]
\centering
\includegraphics[width=6 cm, angle=-90]{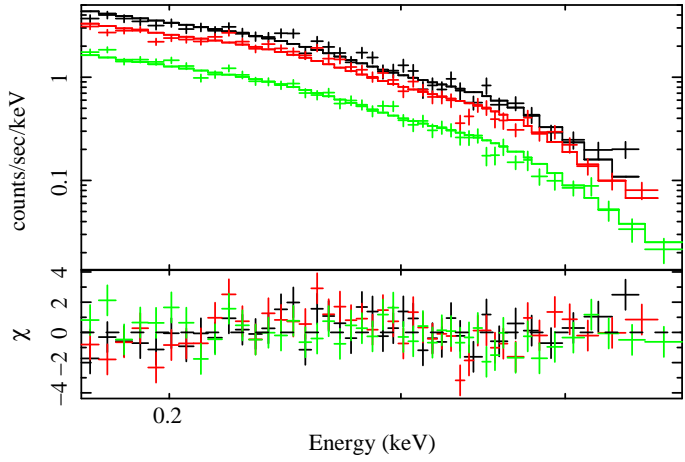}
\caption{\small{Spectra of \sm\ fitted  with a NLTE model atmosphere
with $\log g$  = 6 (see Table \ref{fit_neb_ab}).
Top panel: data points and  model. Bottom panel: residuals from the
model in units of $\sigma$.}}
\label{atm}
\end{figure}

We also explored fits with model atmospheres computed
under non local thermodynamic equilibrium (NLTE) conditions.
A small grid of model atmospheres was calculated with
\emph{TMAP}\footnote{http://astro.uni-tuebingen.de/\raisebox{.3em}{{\tiny $\sim$}}rauch/TMAP/TMAP.html},
the T\"ubingen Model-Atmosphere Package \citep{wer03,rau03b}.
These models, that for a given chemical composition have $\log g$ and $T_\mathrm{eff}$
as free parameters, were converted to the \emph{XSPEC} atable format for spectral fitting.
In the framework of the Virtual Observatory,
the spectral energy distributions (SEDs) of these models are available via the
German Astrophysical Virtual Observatory (\emph{GAVO}) service
\emph{TheoSSA}\footnote{http://vo.ari.uni-heidelberg.de/ssatr-0.01/TrSpectra.jsp?}.

Adopting a  pure H composition, good fits could be obtained for effective temperatures
of (0.7--1.1)$\times10^5$ K,  but with unconstrained values of gravity
in the wide range 5$<$ $\log g$  $<$9 covered by the models.
Equally good fits were given by a He atmosphere with $T_\mathrm{eff}$=(1--1.2)$\times10^5$ K and in this
case the surface gravity could be constrained in the range $\log g$  $\sim$5.5$\pm$0.5.
The best fit effective temperatures  of the H and He NLTE model fits imply source emitting radii of $\sim$1--10 $\rsun$, in order to provide the observed source luminosity. Such radii, coupled to the corresponding $\log g$  values, lead to unrealistically high   values
for the mass of the star
($>$100 $\msun$). We thus conclude that, although the H and He NLTE model atmospheres give formally good fits to the EPIC data, they are physically unacceptable.

We finally considered a NLTE model atmosphere including other elements.
Considering that the limited spectral resolution of our data
does not allow a detailed determination of the elemental composition and abundances,
we restricted the analysis to a single model, based on the results of
optical/UV spectroscopy of the \sm\ nebular emission. We adopted therefore a model
composed of H, He, C, N, and O with the
abundances fixed to the values determined by \citet{lei96}, and reported in
Table \ref{neb_ab}.
The results obtained with this model are summarized
in Table \ref{fit_neb_ab},
where, for three fixed values of $\log g$, we report the interstellar
absorption, effective temperature, and radius of the emission surface derived from the fit
(columns 3, 4,5), as well as the star's mass implied by the $\log g$ and R values (column 6).

In all cases an effective
temperature of $T_\mathrm{eff}\sim1.5\times10^5$ K is obtained.
Although formally the best fit is found for $\log g$ =7, the $\log g$ =6 case (Fig. \ref{atm}) is to be preferred since it is only marginally worse but gives more plausible values of mass ($\sim$1 $\msun$) and radius ($\sim$0.2 $\rsun$).
We checked that the flux in the visible band predicted by this model is
smaller ($\sim$30\%) than the current upper limit  obtained with the
Hubble Space Telescope for the PN central star \citep{vil04}.

\begin{table}[ht!]
\centering
\begin{tabular}{cc}
\noalign{\smallskip}
\hline
\noalign{\smallskip}
\multicolumn{2}{c}{Mass fraction} \\
\noalign{\smallskip} \hline \noalign{\smallskip}
H & $6.808\times10^{-1}$\\
He & $3.176\times10^{-1}$\\
C & $9.104\times10^{-5}$\\
N & $1.061\times10^{-3}$\\
O & $4.205\times10^{-4}$\\
\noalign{\smallskip}
\hline
\end{tabular}
\caption{\small{Element abundances used in the NLTE model atmospheres.}}
\label{neb_ab}
\end{table}

\begin{table*}[!t]
\centering
\begin{tabular}{cccccccc}
\noalign{\smallskip}
 \hline
\noalign{\smallskip}
$\log{g}$&$\chi^{2}_{red}$/dof &N$_H$                                             & T                 & R & M& $F_{X}$ & $F_{bol}$\\
          (cm sec$^{-2}$)&          & ($10^{20}\, \mbox{cm}^{-2}$) &($10^{5}$ K) & ($10^{10}$ cm) & ($10^{33}$ g)& ($10^{-11}$ $\frac{erg}{cm^{2} s}$) &($10^{-10}$ $\frac{erg}{cm^{2} s})$\\
\noalign{\smallskip} \hline \noalign{\smallskip}
 6.0	&1.17/124	& $2.8\pm0.4$ & $1.54\pm0.01$ & $1.3\pm0.1$ & $2.4\pm0.2$ & $2.96$ & $1.51$ \\\noalign{\smallskip}
  6.5	&1.11/124	& $3.2^{+0.3}_{-0.5}$ & $1.54^{+0.01}_{-0.02}$ & $1.5\pm0.1$ & $10.1\pm0.6$ & $3.5$ & $2.0$ \\\noalign{\smallskip}
 7.0	&1.09/124	& $3.0\pm0.5$ & $1.54^{+0.01}_{-0.02}$ & $1.3\pm0.1$ & $25.1^{+2.2}_{-1.3}$ & $2.73$ & $1.56$ \\
\noalign{\smallskip}
\hline
\end{tabular}
\caption{\small{Fit parameters for a H/He/C/N/O NLTE atmosphere;  $F_{X}$ is the unabsorbed source flux in the 0.1-0.5 keV energy band.}}
\label{fit_neb_ab}
\end{table*}

\begin{table*}[ht!]
\centering
\begin{tabular}{lcccc}
\noalign{\smallskip} \hline\noalign{\smallskip}
Name$^{(1)}$ & Rate$^{(2)}$ & L$_{BB}^{(3)}$ & L$_{TB}^{(4)}$ & Observation Date \\
	& ($10^{-3}$ cts s$^{-1}$)	 & ($10^{34}$ erg s$^{-1}$) & ($10^{34}$ erg s$^{-1}$) & \\
\noalign{\smallskip}
\hline\noalign{\smallskip}
SMP SMC 5	 &   3.7 &  67.5 &   2.5 & 2006/03/27 \\
SMP SMC 7	 &   1.1 &  20.5 &   0.8 & 2009/10/20 \\
SMP SMC 8	 &   2.2 &  40.6 &   1.5 & 2009/11/09 \\
SMP SMC 9	 &   1.5 &  27.5 &   1.0 & 2009/10/03 \\
SMP SMC 10	 &   2.9 &  53.5 &   2.0 & 2009/09/27 \\
SMP SMC 12$^*$&   3.2 &  93.9 &   2.7 & 2007/10/28 \\
SMP SMC 13$^*$&   1.7 &  50.6 &   1.5 & 2007/10/28 \\
SMP SMC 14$^*$&   2.6 &  75.0 &   2.2 & 2007/10/28 \\
SMP SMC 16 	 &   2.5 &  45.2 &   1.7 & 2007/06/23 \\
SMP SMC 18 	 &   2.3 &  43.2 &   1.6 & 2007/04/11 \\
SMP SMC 19   &   1.6 &  29.5 &   1.1 & 2006/11/01 \\
SMP SMC 21 	 &   1.7 &  31.4 &   1.2 & 2009/10/09 \\
SMP SMC 23 	 &   1.9 &  35.0 &   1.3 & 2009/10/11 \\
SMP SMC 25 	 &   $3.2\pm0.9$ &  58.5 &   2.2 & 2007/06/04 \\
SMP SMC 27 	 &   1.2 &  22.9 &   0.9 & 2009/06/29 \\
J2 	 &   1.8 &  33.9 &   1.3 & 2009/10/03 \\
J4$^*$   &   3.5 & 102.3 &   3.0 & 2007/10/28 \\
J5       &   2.8 &  51.6 &   1.9 & 2009/10/03 \\
J9 	 &   3.0 &  56.2 &   2.1 & 2007/04/11 \\
J10 	 &   1.9 &  34.7 &   1.3 & 2009/09/27 \\
J12 	 &   1.8 &  33.7 &   1.3 & 2009/09/27 \\
J13      &   7.0 & 129.1 &   4.8 & 2007/06/23 \\
J15      &   2.3 &  42.4 &   1.6 & 2007/06/23 \\
J16      &   2.5 &  46.1 &   1.7 & 2007/06/23 \\
J17 	 &   2.1 &  39.3 &   1.5 & 2007/04/11 \\
J18 	 &   2.0 &  36.7 &   1.4 & 2006/11/01 \\
J21 	 &   1.6 &  28.6 &   1.1 & 2009/11/04 \\
J22 	 &   3.2 &  58.5 &   2.2 & 2007/06/23 \\
J23 	 &  12.7 & 234.2 &   8.7 & 2009/10/11 \\
J24 	 &   4.4 &  80.8 &   3.0 & 2009/10/11 \\
J25 	 &   1.7 &  30.8 &   1.1 & 2009/10/11 \\
J27 	 &   2.4 &  43.5 &   1.6 & 2009/10/11 \\
MG8 	 &   1.6 &  29.1 &   1.1 & 2009/09/27 \\
MG9 	 &   3.6 &  66.9 &   2.5 & 2007/06/06 \\
MG10 	 &   3.2 &  59.4 &   2.2 & 2006/10/03 \\
MG11 	 &   1.6 &  29.1 &   1.1 & 2009/11/30 \\
MA14 	 &   2.1 &  38.0 &   1.4 & 2009/10/20 \\
MA44 	 &   1.8 &  33.0 &   1.2 & 2009/10/18 \\
MA406$^*$&   3.0 &  86.9 &   2.5 & 2007/10/28 \\
MA891 	 &   4.0 &  74.5 &   2.8 & 2002/03/30 \\
MA1357 	 &   4.2 &  76.9 &   2.9 & 2009/10/16 \\
MA1714 	 &   1.6 &  29.0 &   1.1 & 2009/11/16 \\
MA1762 	 &   2.6 &  47.4 &   1.8 & 2009/11/19 \\
M7 	 &   1.9 &  36.0 &   1.3 & 2009/06/29 \\
JD29 	 &   1.7 &  31.5 &   1.2 & 2009/10/18 \\
JD51 	 &   2.8 &  51.8 &   1.9 & 2009/10/09 \\
JD53 	 &   3.5 &  64.2 &   2.4 & 2007/06/06 \\
JD57 	 &   1.3 &  24.2 &   0.9 & 2009/09/25 \\
\noalign{\smallskip} \hline \noalign{\smallskip}
\end{tabular}
\caption{\small{$3\sigma$ upper limits for planetary nebulae in the
SMC observed with \textit{XMM-Newton} (Except for SMP SMC 25 for which we report the detected count rate). The sources marked by a star were observed with the medium optical filter, all the others with the thin one.
(1) The names indicate the following references:
       SMP = \citet{san78},
	   J = \citet{jac80}, 
       MG = \citet{mor85}, 
	   MA = \citet{mey93}, 
	    M = \citet{mor95}, 
      JD = \citet{jac02}; 
(2) 0.1-1 keV count rate in the EPIC pn camera;
(3) unabsorbed 0.1-1 keV luminosity for a blackbody
spectrum with kT=20 eV,  $N_H=5\times10^{20}$ cm$^{-2}$, and d=60 kpc;
(4) as (3) but for a thermal bremsstrahlung with kT=100 eV.
}}
\label{PN_ul}
\end{table*}

\subsection{Search for X-ray emission from other SMC planetary nebulae}

\begin{figure}[htbp]
\centering
\includegraphics[width=9 cm]{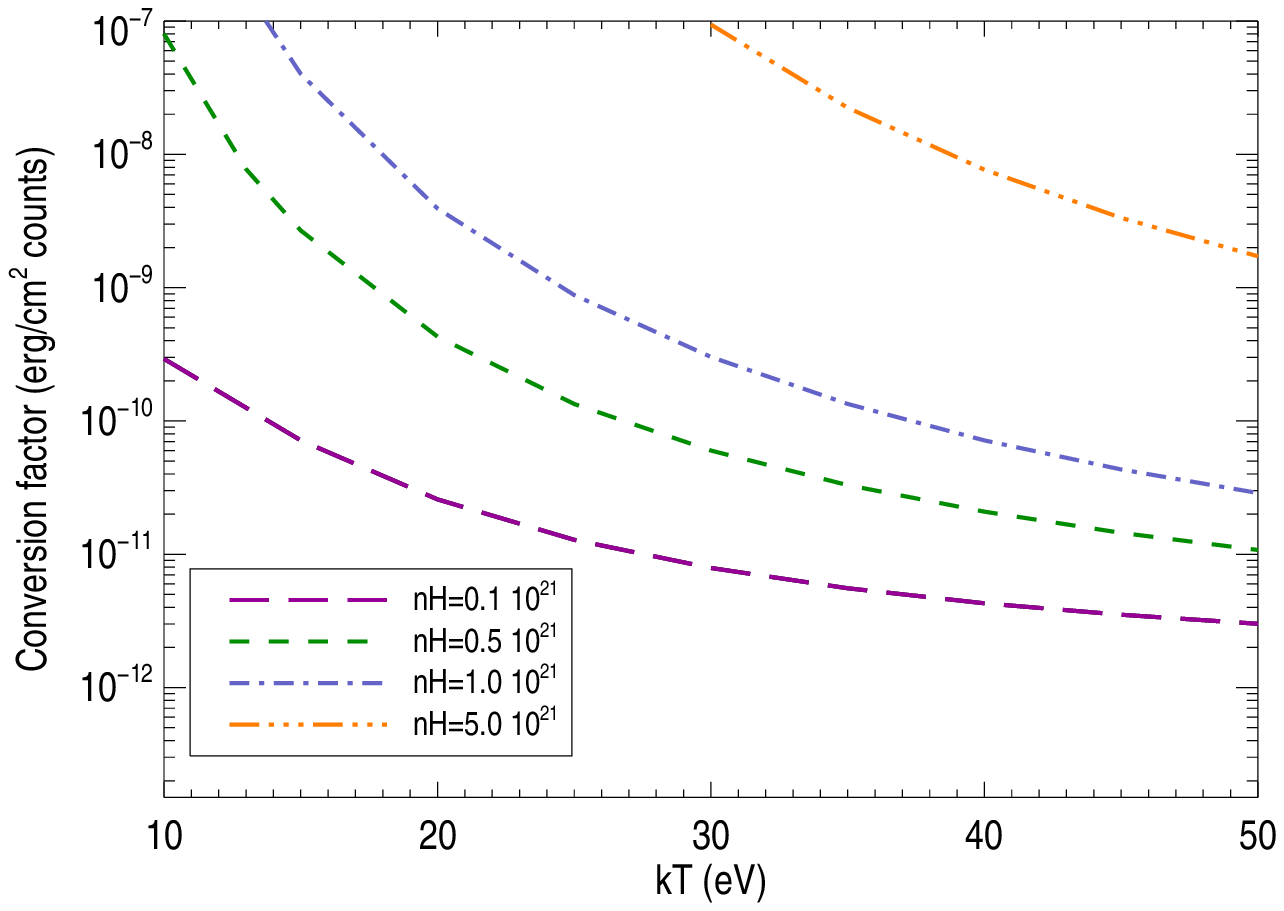}
\includegraphics[width=9 cm]{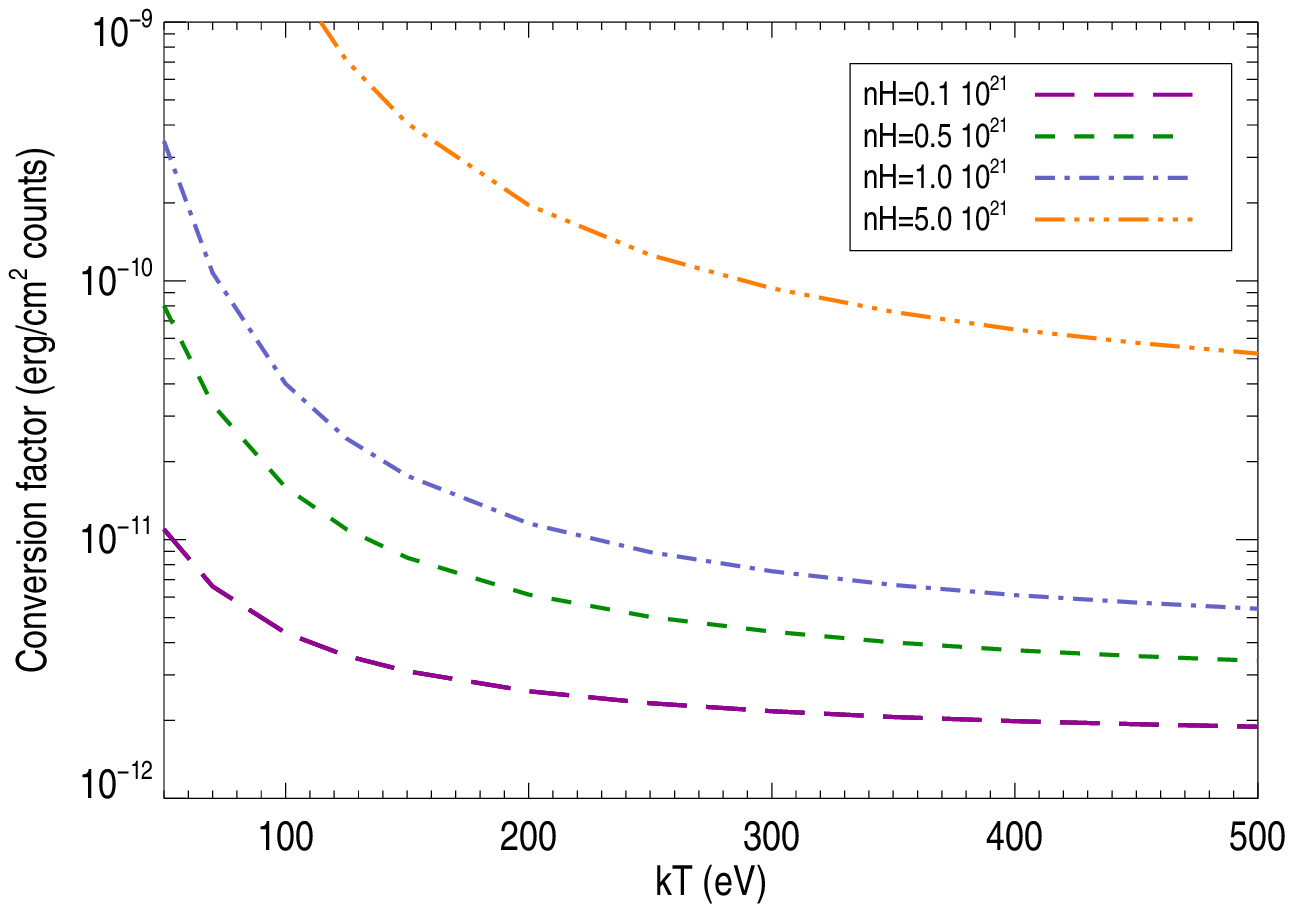}
\caption{\small{Conversion factors from PN count rate to unabsorbed flux in the 0.1-1 keV energy band. Top panel: blackbody model, bottom panel: bremsstrahlung model.
Both plots refer to observations carried out with the thin optical filter.}}
\label{cf}
\end{figure}

Based on several compilations (see references in  Table \ref{PN_ul})
we  selected all the  PNe in the SMC that were observed by \xmm\ in our survey
or in other observations in the public archive.
Using the data from the EPIC pn camera,
we searched for X-ray emission from these objects in the 0.1-1 keV energy band applying
the SAS maximum likelihood  source detection task \textsc{emldetect}.
with   the threshold parameter \textit{mlmin} set to the value 6.
For PNe observed in more than one pointing, typically with different off-axis
angles and net exposure times, we used the data giving the best sensitivity at the source
position.
With the exception of  SMP SMC 25 (see below), none of the sources in our sample was significantly detected.
We report in Table \ref{PN_ul} the upper limits on their 0.1--1 keV count rates,
that take into account the vignetting correction
due to the different off-axis angles.
The conversion of these values to physical units requires
some assumption on the source spectra.
Conversion factors from count rate to flux, for different values of temperature and hydrogen column density, are plotted in Fig. \ref{cf} for the case of thin optical filter (that was used in most observations). As representative examples, we report in Table \ref{PN_ul} the luminosity upper limits  for a blackbody model with $kT=20$ eV and for  a bremsstrahlung plasma model with $kT=100$ eV (in both cases $N_H=5\times10^{20}$ cm$^{-2}$). For comparison, the corresponding values for \sm\ would be L$_{BB}$=1.9$\times10^{38}$  erg s$^{-1}$ and
L$_{TB}$=2.2$\times10^{36}$  erg s$^{-1}$.

The detection of  SMP SMC 25 with \xmm\ confirms the association of the low luminosity
SSS source RX J0059.6--7138, discovered with ROSAT \citep{kah99},
with this planetary nebula, also known as LIN357 \citep{mey93}.
The coordinates derived with EPIC (R.A.= 0$^h$ 59$^m$ 40.3$^s$, Dec.= --71$^{\circ}$ 38$'$ 17$''$, with a 1$\sigma$ uncertainty
of 2.5$''$) are consistent with the optical position of SMP SMC 25. For comparison, the
90\% c.l. error radius  obtained for RX J0059.6$-$7138  with ROSAT was  14.5$''$ \citep{hab00}. The EPIC and ROSAT count rates, considering the uncertainties
due to the small statistics and poorly constrained spectrum, are consistent with a constant
source flux.

\section{Discussion}

Due to its location in the SMC,
\sm\ has angular dimensions too small to be spatially resolved with EPIC:
0.71$''\times$0.54$''$  \citep{sta03,vas98}.
However, the very soft spectrum and high luminosity derived with our spectral
analysis imply that its X-ray emission  originates from the central star, without
any significant contribution from the diffuse gas.
The  blackbody fit gives a bolometric
luminosity L$_{bol}$$\sim8\times10^{37}$ erg s$^{-1}$,
while the more realistic atmosphere model with the nebular abundances yields
L$_{bol}=6\times10^{37}$ erg s$^{-1}$.
Such a value is orders of magnitude higher than the luminosity that can be
produced in the surrounding shock-heated gas.
The same conclusion was reached by \citet{wan91}, who, due to the lack of
adequate X-ray spectral information, assumed a blackbody
with T=3$\times$10$^5$ K.

Previous spectral analysis of \sm\ were carried out with the Einstein Observatory
by \citet{bro94} and with ROSAT by \citet{kah94}. A  blackbody model was used in
both cases, yielding best fit parameters consistent with, but much less constrained, than
those derived here in our analysis.
The comparison of these data, spanning almost 30 years, does not give any
evidence for long term variability of the source luminosity and/or spectrum.
\citet{hei94} fitted the ROSAT spectrum of \sm\ with models
of H-rich and He-rich  white dwarf  atmospheres computed with the LTE assumption.
They found that these models,  appropriate for H- or He-burning accreting
white dwarfs, yield a bolometric luminosity of $\sim2\times10^{37}$ erg s$^{-1}$,
more than one order of magnitude smaller than the super-Eddington value
implied by the blackbody fit to the same data.
However, the lack of variability and the upper limit (V$>$20.7) on the
optical counterpart \citep{vil04}, do not favor a binary nature,
contrary to the case of other SSS.
The good fit provided by our NLTE model supports instead the interpretation of \sm\ as
a single, very hot star on its way to become a relatively massive  ($\sim$1 $\msun$)
white dwarf.
Another hint for a high mass star  comes from the high N/O ratio
(see Table \ref{neb_ab})  consistent with a Type I PN, implying a
massive progenitor (see, e.g. \citealt{sta07}).

\begin{figure}[htbp]
\hspace{-15 pt}
\includegraphics[width=9.5 cm]{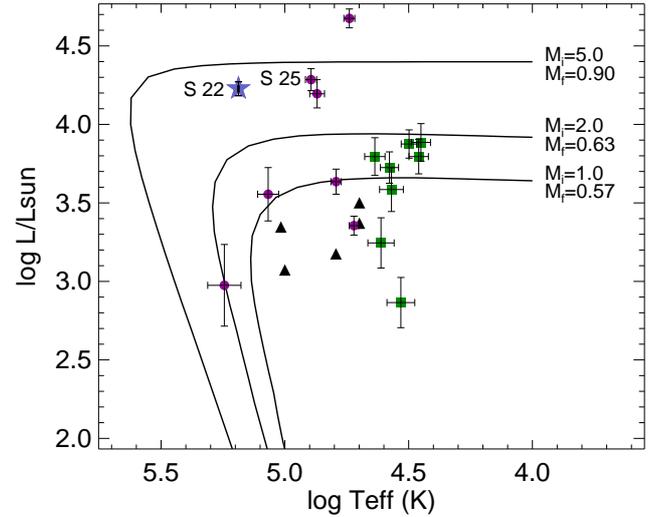}
\caption{\small{H-R diagram for central stars of PNe in the SMC. Black triangles are from \citet{all87}, purple circles and green squares from \citet{vil04} (Zanstra He and H temperature respectively). The blue star represents the nucleus of \sm . Evolutionary tracks for SMC metallicity are from \citet{vas94}.   }}
\label{HR}
\end{figure}

It is interesting to compare the temperature of the central star derived from our
X-ray spectral analysis (T$\sim1.5\times10^5$ K) with the values obtained from the
modeling of nebular emission lines observed at optical/UV wavelengths.
Based on the Zanstra method, which assumes a blackbody spectrum for the ionizing
radiation from the central star \citep{zan27},
\citet{vil04} derived T$_{HeII}=(1.222\pm0.145)\times10^5$ K
and T$_{H}=(0.77\pm0.18)\times10^5$ K, using the He II 4686\AA~ and H$_{\beta}$ lines, respectively. A temperature T=1.15$\times10^5$ K was instead obtained
from IUE spectroscopy, by adopting
NLTE model atmospheres \citep{all87}. These values are in good agreement
with our results,  considering the unavoidable uncertainties intrinsic in all
these model-dependent estimates.

The X-ray detection of \sm\ is not surprising, considering that
all the studies based on optical/UV data  indicate
that this PN hosts one of the hottest
central stars.
On the other hand, the same studies also show the presence in the SMC of other PNe with similarly high temperature, as   illustrated by the
H-R diagram plotted in Fig. \ref{HR}.
Based on this plot we would expect other objects with a soft X-ray emission
comparable to, or slightly lower than, that of \sm .
Instead, our results (Table \ref{PN_ul}) indicate that
most of these PNe are   more than two
orders of magnitude less luminous in the X-ray band than \sm .
This applies also to SMP SMC 25, which has a
Zanstra temperature  about a factor two lower than that of
\sm\  \citep{vil04}, accounting for the large   difference of soft X-ray flux   between these two PNe.
Thus, the  X-ray faintness of
SMC PNe with hot central stars is likely due to the combination
of absorption in the nebula and of the strong dependence of the flux in the EPIC
soft X-ray band on the temperature (see Fig. \ref{cf}).
The small value of absorption derived from our fits of \sm\ is consistent with the Galactic
value in the SMC direction, indicating only little or no intrinsic absorption in the nebula
itself. The high temperature and luminosity of \sm\ are also the likely explanation
for the absence of dust related features in the Spitzer infrared data  \citep{ber09}.
\sm\ is one of the really energetic PNe with high excitation lines
([O IV] and [Ne V]) as the IR Spitzer spectra show. However, it is not a
unique object, as several of the SMC PNe in the \citet{ber09}  study show these lines.
This also supports our argument that the
absorption may play a major role in hiding other massive PNe across the
whole SMC. Recently, \citet{fil09a} and \citet{pay08}
detected four radio-continuum PNe in the SMC which also appear to have very
large and massive central stars. They even tentatively called them ''Super
PNe'' implying their extraordinary nature. We have searched all available
radio-continuum images and catalogues at various radio frequencies
and found no emission from \sm\ and SMP SMC 25 down to a  3$\sigma$ limit of 0.3 mJy \citep{fil97,fil98,fil02}.

PNe in our Galaxy have been observed with X-ray luminosity reaching at most $\sim$10$^{32}$ erg s$^{-1}$ \citep{kas07}, but the detection of Super Soft
Sources such as \sm\ and SMP SMC 25
in the Galactic plane direction is hampered by  interstellar absorption.
Not surprisingly, the PN  most resembling \sm\ in its X-ray properties
has been found in the Large Magellanic Cloud: \xmm\  observations reported by
\citet{kah08}
showed that SMP LMC 29 has a soft  X-ray spectrum well fit by a blackbody
with T in the range (3--6)$\times10^5$ K
and a bolometric luminosity of (0.1--30)$\times10^{36}$   erg s$^{-1}$.

\section{Conclusions}

The first X-ray observations of the SMC planetary nebula \sm\ obtained with a modern
high-throughput satellite have allowed  us to study its X-ray emission
with unprecedented statistics. No evidence for a binary nature,
such as long or short term variability, as seen in other SSS was found.
It is remarkable that, despite different spectral models can fit the
data, a self-consistent picture in terms of temperature, mass and radius of the
central star can be obtained with a NLTE model atmosphere with the same elemental
abundances seen in the nebula.
The inferred mass for the central star, of the order of 1 $\msun$,   implies that
\sm\ is the descendent  of a relatively massive progenitor (see Fig. \ref{HR}).
This may explain its exceptional luminosity,   as well as the apparent rarity
of such objects that evolve very quickly toward the cooling white dwarfs sequence.

\begin{acknowledgements}

This work is based on  observations obtained with \xmm, an ESA
science mission  with instruments  and  contributions directly
funded by  ESA Member States  and NASA.  The  \xmm\ data analysis
is supported  by the Italian  Space  Agency  (ASI). TR is supported
by the German Aerospace Center (DLR) under grant 05\, OR 0806.
\end{acknowledgements}

\bibliographystyle{aa}
\bibliography{biblio_SMP22}

\end{document}